\documentclass[epj]{webofc}
\usepackage[utf8]{inputenc}
\usepackage[varg]{txfonts}   % Web of Conferences font
\usepackage{booktabs}
\usepackage{xcolor}
\usepackage{amsmath,amssymb,epsfig,bm,pifont}
\usepackage{graphicx}
\usepackage{epstopdf}
\usepackage{physics}
\usepackage{subcaption}
\usepackage{wrapfig}
\definecolor{darkred}{rgb}{0.4,0.0,0.0}
\definecolor{darkgreen}{rgb}{0.0,0.4,0.0}
\definecolor{darkblue}{rgb}{0.0,0.0,0.4}
\usepackage[bookmarks,linktocpage,colorlinks,
    linkcolor = darkred,
    urlcolor  = darkblue,
    citecolor = darkgreen]{hyperref}
\wocname{EPJ Web of Conferences}
\woctitle{Lattice2017}
\def\0#1#2{\frac{#1}{#2}}

\def\s0#1#2{\mbox{\small{$ \frac{#1}{#2} $}}}

\newcommand{\beq}{\begin{equation}}
\newcommand{\eeq}{\end{equation}}
\newcommand{\bea}{\begin{eqnarray}}
\newcommand{\eea}{\end{eqnarray}}
\begin{document}
%%%%%%%%%%%%%%%%%%%%%%%%%%%%%%%%%%%%%%%%%%%%%%%%%%%%%%%%%%%%%%%%%%%%%%%%%%%%%
%
\selectlanguage{english}
%----------------------------------------------------------------------------
\title{%
Dynamics of entanglement entropy of interacting fermions in a 1D driven harmonic trap
}
%----------------------------------------------------------------------------
\author{%
\firstname{Joshua R.} \lastname{McKenney}\inst{1}\fnsep\thanks{Speaker, \email{joshmck@unc.edu}} \and
\firstname{William J.} \lastname{Porter}\inst{1} \and
\firstname{Joaqu{\' \i}n E.}  \lastname{Drut}\inst{1}
% etc.
}
%----------------------------------------------------------------------------
\institute{%
Department of Physics and Astronomy, University of North Carolina, Chapel Hill, NC, 27599, USA
}
%----------------------------------------------------------------------------
\abstract{%
Following up on a recent analysis of two cold atoms in a time-dependent harmonic trap in one dimension, we explore the 
entanglement entropy of two and three fermions in the same situation when driven through a parametric resonance. 
We find that the presence of such a resonance in the two-particle system leaves a clear imprint on the entanglement 
entropy. We show how the signal is modified by attractive and repulsive contact interactions, and how it remains present for 
the three-particle system. Additionaly, we extend the work of recent experiments to demonstrate how restricting observation to a 
limited subsystem gives rise to locally thermal behavior.
}
%----------------------------------------------------------------------------
\maketitle
%----------------------------------------------------------------------------
\section{Introduction}\label{intro}

The link between the unitary evolution of microscopic quantum systems and the ergodic dynamics of macroscopic classical systems has been under investigation since the earliest days of quantum statistical mechanics. The fundamental question can be stated as, ``How does a quantum state, which remains pure under action of the time-evolution operator, give rise to thermal behavior?''  A prominent theory to address this issue is the eigenstate thermalization hypothesis, which is closely related to quantum chaos and entanglement (see e.g.~\cite{PhysRevE.50.888,DAlessio:2016rwt,Gogolin}. More recently, experiments have directly shown local thermal behavior in globally pure states~\cite{Kaufman794}.

In this work, we investigate local thermalization by exploring the time-dependent dynamics of entanglement in small systems of spin-$1/2$ particles in a one-dimensional harmonic trap. 
Specifically, we consider the two-body problem with one particle per species, which we will call the 1+1 case, 
and the three-body problem of two particles of one species and one of the other, i.e. 2+1 particles. We focus on a situation recently analyzed 
in Ref.~\cite{Ebert:2015gko}, where it was found that the interacting and noninteracting two-body problems can display dramatically different behaviors when 
driven through a parametric resonance. Specifically, we consider dynamics governed by a Hamiltonian
\beq
\label{Eq:Hamiltonian}
\hat{H}= \hat T + \hat V_I + \hat V_\text{ext},
\eeq
where the kinetic and interaction energy operators are time-independent and given by
\beq
\hat T =
-\frac{1}{2}
\sum_{\sigma=\uparrow,\downarrow}
\int d^{}x\  
\hat{\psi}_{\sigma}^{\dagger}(x)\frac{d^2}{dx^2}\hat{\psi}_{\sigma} (x),
\eeq
and
\beq
\hat V_I = 
-g
\int d^{}x\ 
\hat{n}_\uparrow (x) \hat{n}_\downarrow (x),
\eeq
respectively, whereas the external potential is
\beq
\hat V_\text{ext} = 
\frac{\omega^2(t)}{2}
 \sum_{\sigma=\uparrow,\downarrow}
\int d^{}x \;
x^2 \hat{n}_\sigma (x),
\eeq
where 
%
%\beq
$\omega^2(t) = \omega_0^2\left(1+\alpha \sin\left(t/T\right)\right)$,
%\eeq
%
for a fixed time-independent frequency $\omega_0$ and parameters $\alpha, T$ to be
explored; note $\alpha$ is dimensionless, but $T$ has units of time, i.e. inverse energy. This 
form was chosen because it follows the protocol of Ref.~\cite{Ebert:2015gko}.
As usual, the operators $\hat{\psi}^{\dagger}_{s}, \hat{\psi}^{}_{s}$ used above are the creation and annihilation operators for particles 
of spin $s=\uparrow,\downarrow$ and $\hat{n}^{}_{s}$ is the corresponding particle density operator. The coupling $g$ is directly related 
to the scattering length $a_0$ via $g = 2/a_0$, and we set $m=\hbar=1$.

The work of Ref.~\cite{Ebert:2015gko} compared the behavior of different cases with $g \leq 0$, i.e. either noninteracting or repulsively interacting
(note our sign convention). Here, we reproduce part of that analysis as a way to check our calculations and include the attractively interacting 
case as well. We further extend that work by studying the behavior of the R\'enyi entanglement entropy (defined next)
as a function of sub-system size and for different coupling strengths and driving parameters. To provide a more intuitive measure of thermal behavior, we present our results as the fraction of the Hilbert space required to describe the system.

%%%%%%%%%%%%%%%%%%%%%%%%%%%%
\subsection{Entropy}
Since the seminal work of Shannon in the 1940s~\cite{BLTJ:BLTJ1338}, entropy has been widely used as a precise means of quantifying information. 
Defined for a discrete random variable $X$ with possible values $x$, the Shannon entropy $S_S$ takes the form
\beq
S_S(X) = -\sum_x p(x)\log{p(x)}, 
\eeq
where the summation becomes an integral for continuous $X$. The essential question that this quantity answers is roughly, 
``How much additional information is needed to unambiguously specify a
particular configuration?'' Thus for states that are fully known, the entropy is zero, while for states with high multiplicity, the entropy is large. 
In this sense, the Shannon entropy parallels the combinatorial entropy of the microcanonical ensemble, with its macrostates and microstates. 
For uniform probability distributions $p(x)$, the Shannon entropy reduces to the Boltzmann entropy;
on the other hand, if $p(x)$ is defined by Boltzmann weights and the partition function, the Shannon entropy gives the same result as the canonical ensemble for the entropy of the ideal gas 
(under the usual assumptions of indistinguishability and phase space in units of Planck's constant $h$). 

The quantum mechanical analog of the Shannon entropy is the von Neumann entropy,
\beq
S_{vN} = -\Tr\left[\hat{\rho} \log{\hat{\rho}} \right],
\eeq
where $\hat{\rho}$ is the density operator. The ``variable'' now under consideration is the space of states accessible to a quantum system, and the missing information quantified by this entropy is
the amount needed to distinguish which particular state the system is in. The von Neumann entropy may be considered as a special case ($n\rightarrow 1$) of the more general $n$-th order
R\'enyi entropies,
\beq
S_n = \frac{1}{1-n} \log{\Tr\hat{\rho}^n}.
\eeq
Of particular physical interest is the second order R\'enyi entropy,
\beq
S_2 = -\log{\Tr\hat{\rho}^2},
\eeq
as $\Tr\hat{\rho}^2$ reflects the purity of a quantum state, equal to 1 for a pure state and less than 1 otherwise. Correspondingly, $S_2$ grows with increased mixing of states. For a maximally mixed state, $S_2 = \log{V_{\mathcal{H}}} = \log\left(\text{rank}\ \hat{\rho}\right)$, where $V_{\mathcal{H}}$ is the volume of the Hilbert space. Accordingly, we identify $\exp(S_2)$ as the weighted number of
states contributing to $\hat{\rho}$ and introduce the fraction of the Hilbert space occupied, $\exp(S_2)/V_{\mathcal{H}}$, as a normalized measure valid for any state. When this quantity approaches unity, all states become equally likely to be occupied, in analogy to the infinite-temperature limit.

%%%%%%%%%%%%%%%%%%%%%%%%%%%%%%%%%%%%%
\subsection{Entropy of entanglement and reduced density matrices}

To characterize spatial entanglement, we divide the 1D region on which the full Hilbert space $\mathcal{H}$ exists into two subregions, 
$A$ and its complement, $\bar{A}$. The full space $\mathcal{H}$ may be written as a direct product space of the Hilbert spaces of the subregions as
\beq
\mathcal{H} = \mathcal{H}_A \otimes \mathcal{H}_{\bar{A}}.
\eeq
Since the subsystem $A$ is our focus and its complement $\bar{A}$ is the ``environment,'' we have access only to the density matrix of A, $\hat{\rho}_A$, which is the reduced density matrix
of the full system,
\beq
\hat{\rho}_A = \Tr_{\mathcal{H}_{\bar{A}}} \hat{\rho},
\eeq
where the trace is over the states of $\mathcal{H}_{\bar{A}}$ in the position-space basis.

Once $\hat \rho_A$ is known, the R\'enyi and von Neumann entanglement entropies can be computed via
\beq
S_{vN} = -\Tr\left[\hat{\rho}_A \log{\hat{\rho}_A} \right],
\eeq
and
\beq
S_n = \frac{1}{1-n} \log{\Tr\hat{\rho}_A^n}.
\eeq
Both entropies naturally vanish when $A$ is an empty set or the whole system, as then $\hat{\rho}_A$ corresponds 
to a single state (which is trivially factorizable).

%%%%%%%%%%%%%%%%%%%%%%%%%%%
\subsection{Computational method, lattice, scales}

To study the small systems we are concerned with here, we employ the non-stochastic method of Ref.~\cite{Porter:2016yaf} to build the
transfer matrix 
\beq
\hat{\mathcal T} = \exp(-\tau \hat H).
\eeq
\begin{figure*}[thb]
\centering%\captionsetup{format=plain,justification=centerlast}
\includegraphics[width=0.45\columnwidth]{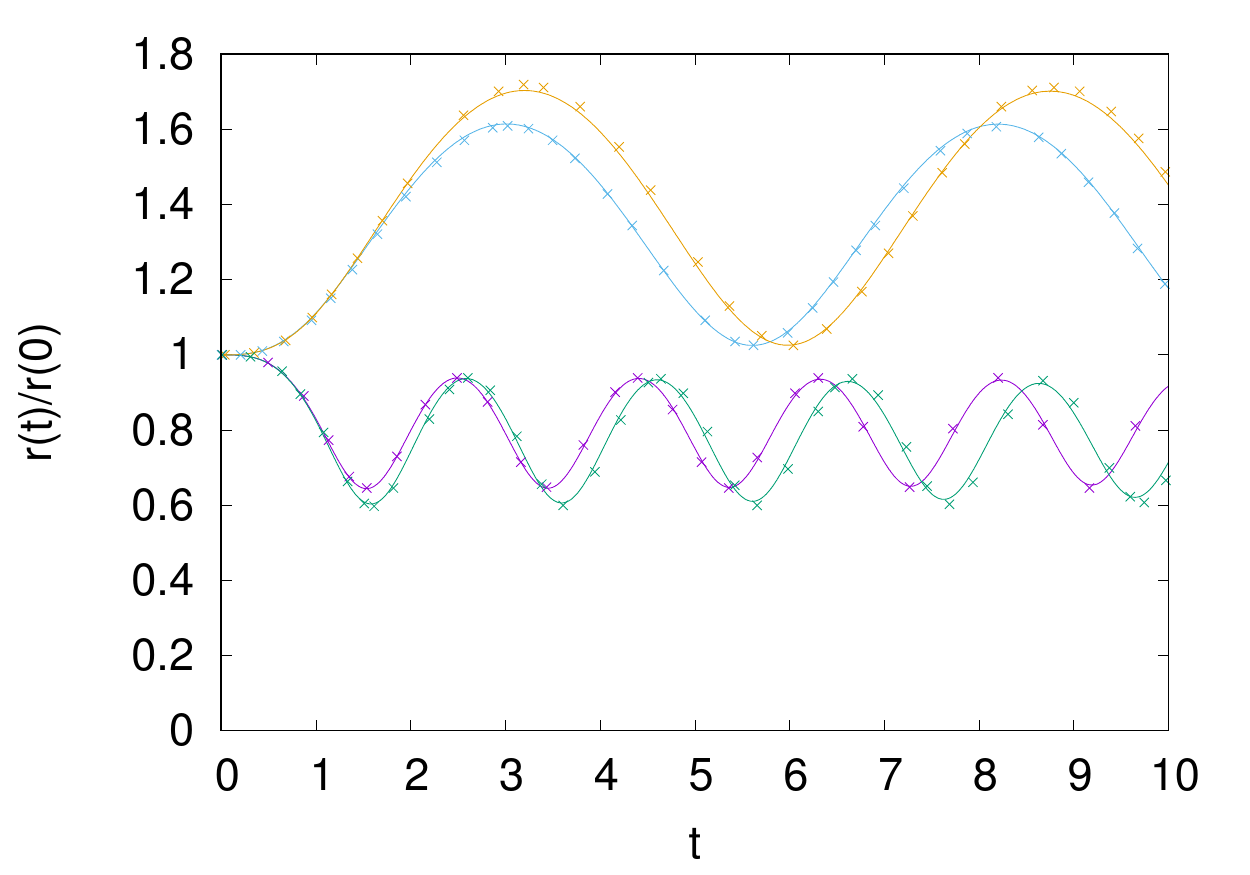}
\includegraphics[width=0.45\columnwidth]{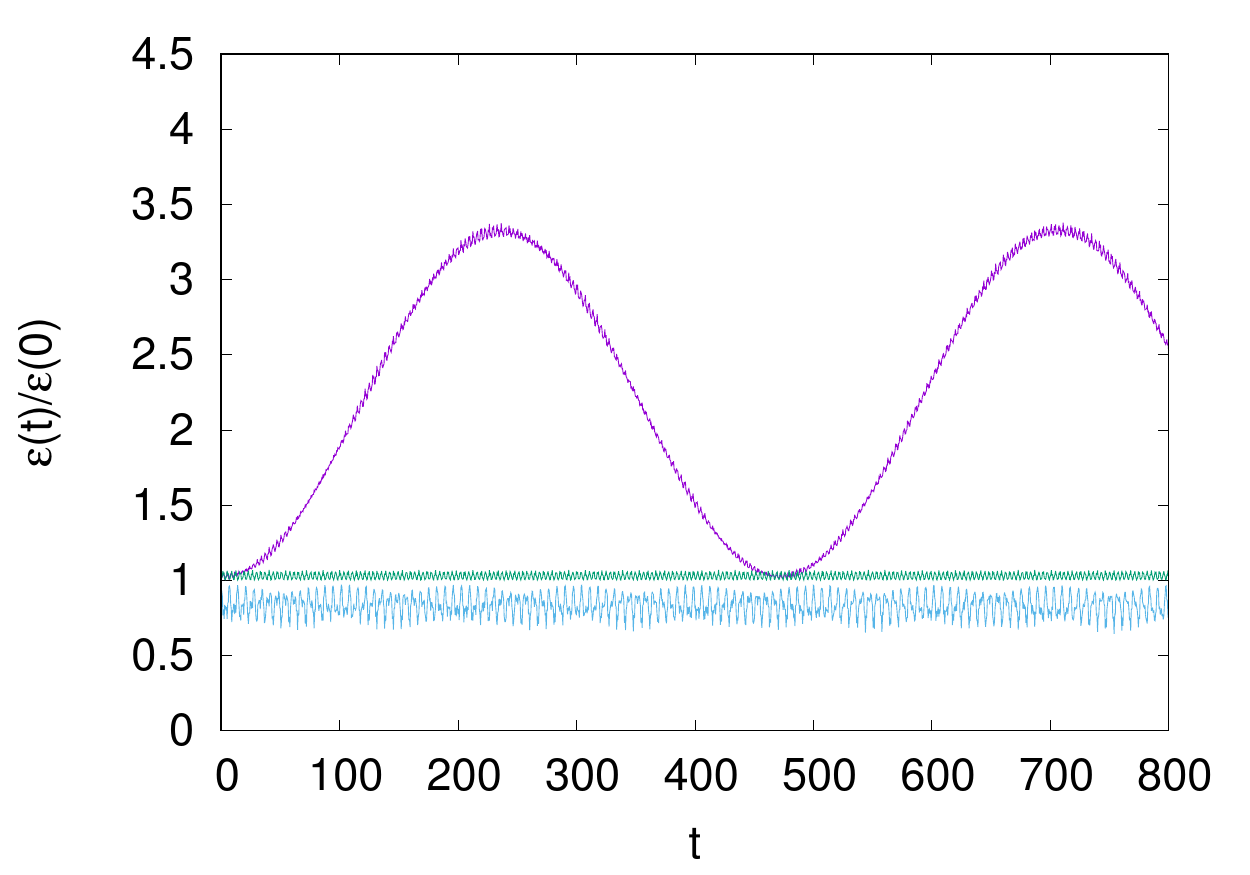}
\caption{\label{Fig:DarmRMS} 
\textbf{(Left)} Root-mean-square separation of two-body system as a function of time for exponential quenches in repulsive (yellow, green) and noninteracting (blue, purple) cases. 
Overlaid crosses `$\times$' mark samples of data from Ref.~\cite{Ebert:2015gko} (see for more details). \textbf{(Right)} Energy as function of time for $\alpha=0.5,\ T=1/3.80$ for attractive [$g=2$ (blue)], noninteracting [$g=0$ (green)], 
and repulsive [$g=-2$ (purple)] cases.
}
\end{figure*}
\begin{figure*}[thb]
\centering\captionsetup{format=plain,justification=centerlast}
	\includegraphics[width=0.32\columnwidth]{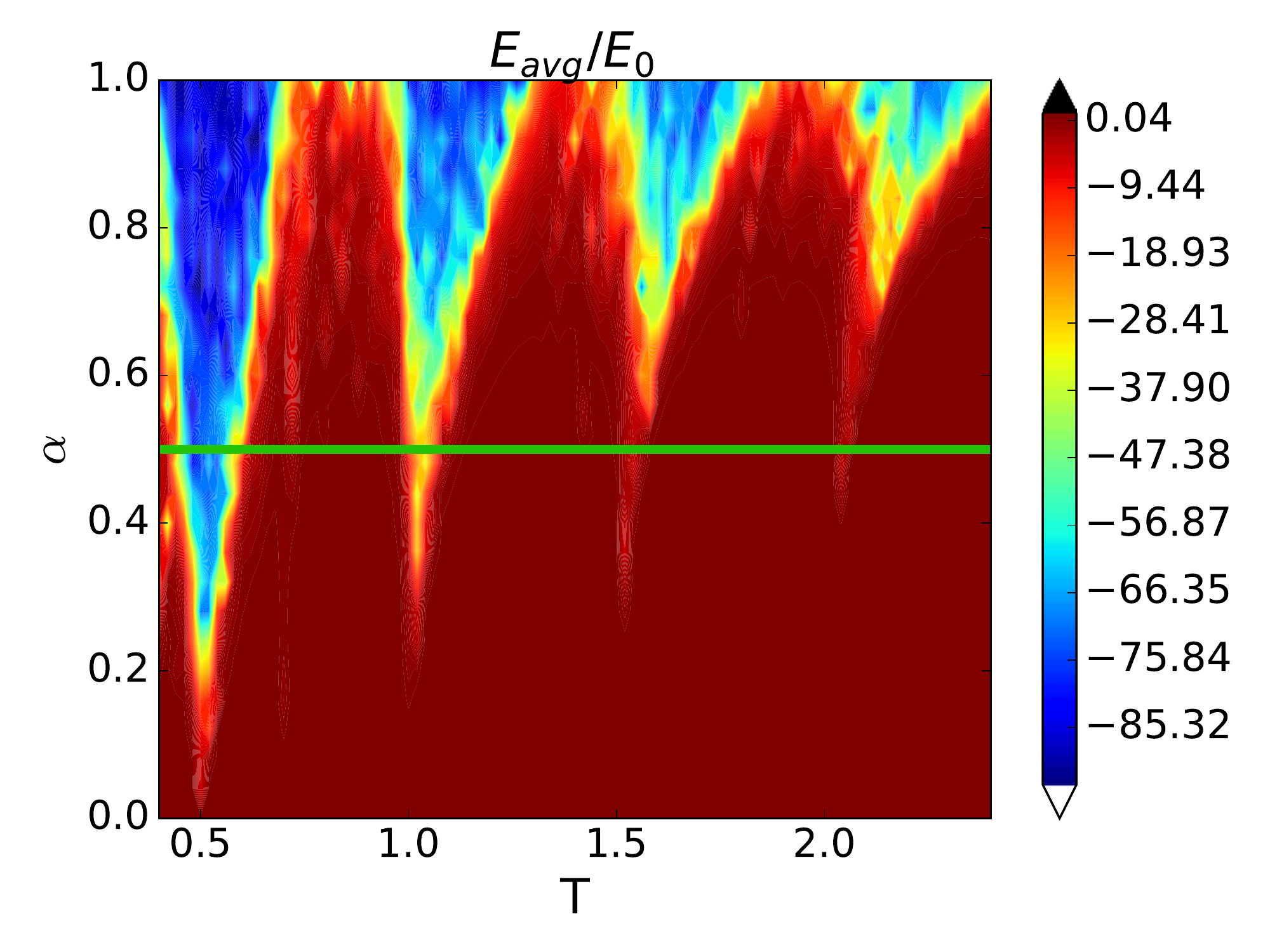}
	\includegraphics[width=0.32\columnwidth]{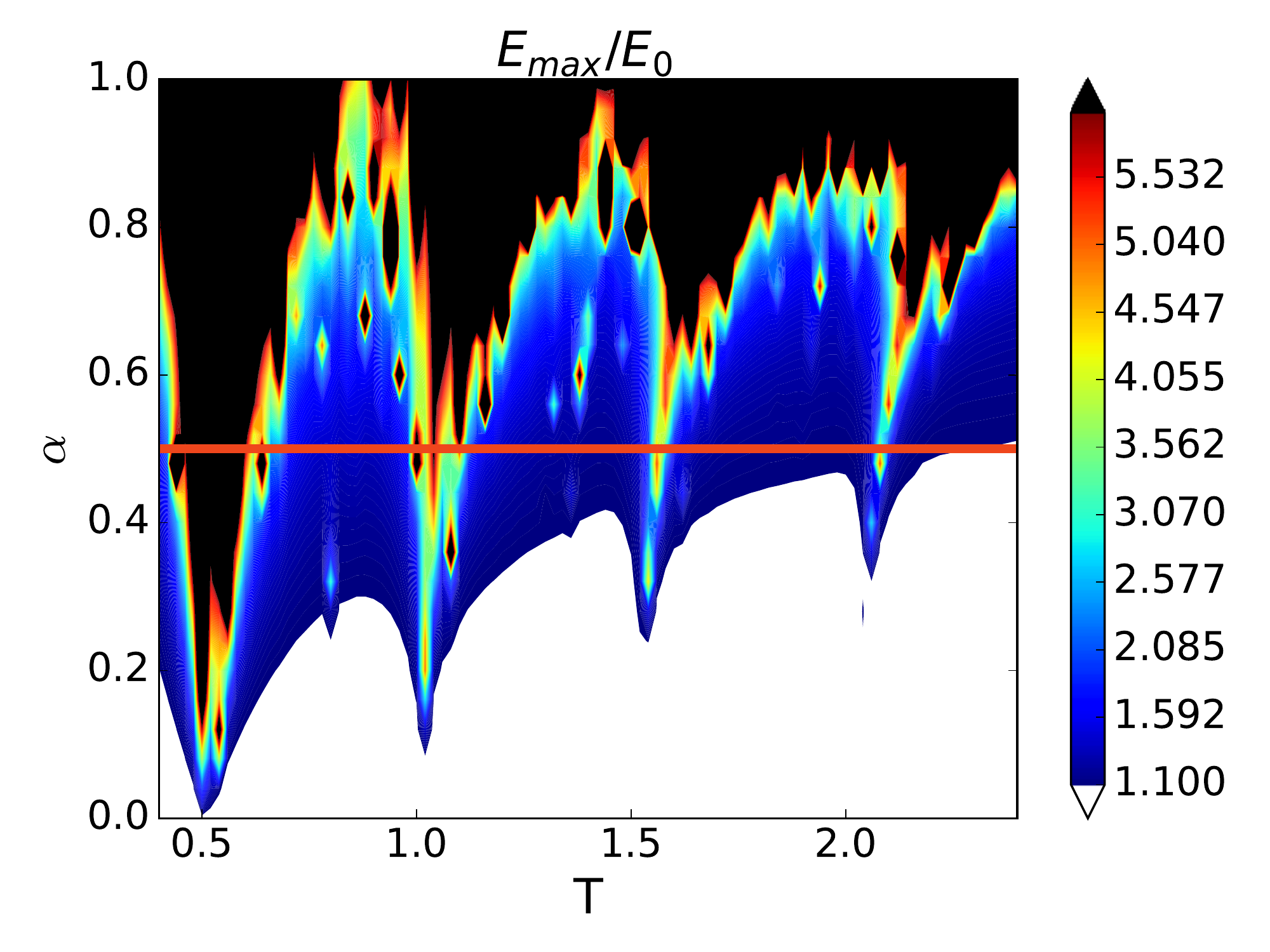}
	\includegraphics[width=0.32\columnwidth]{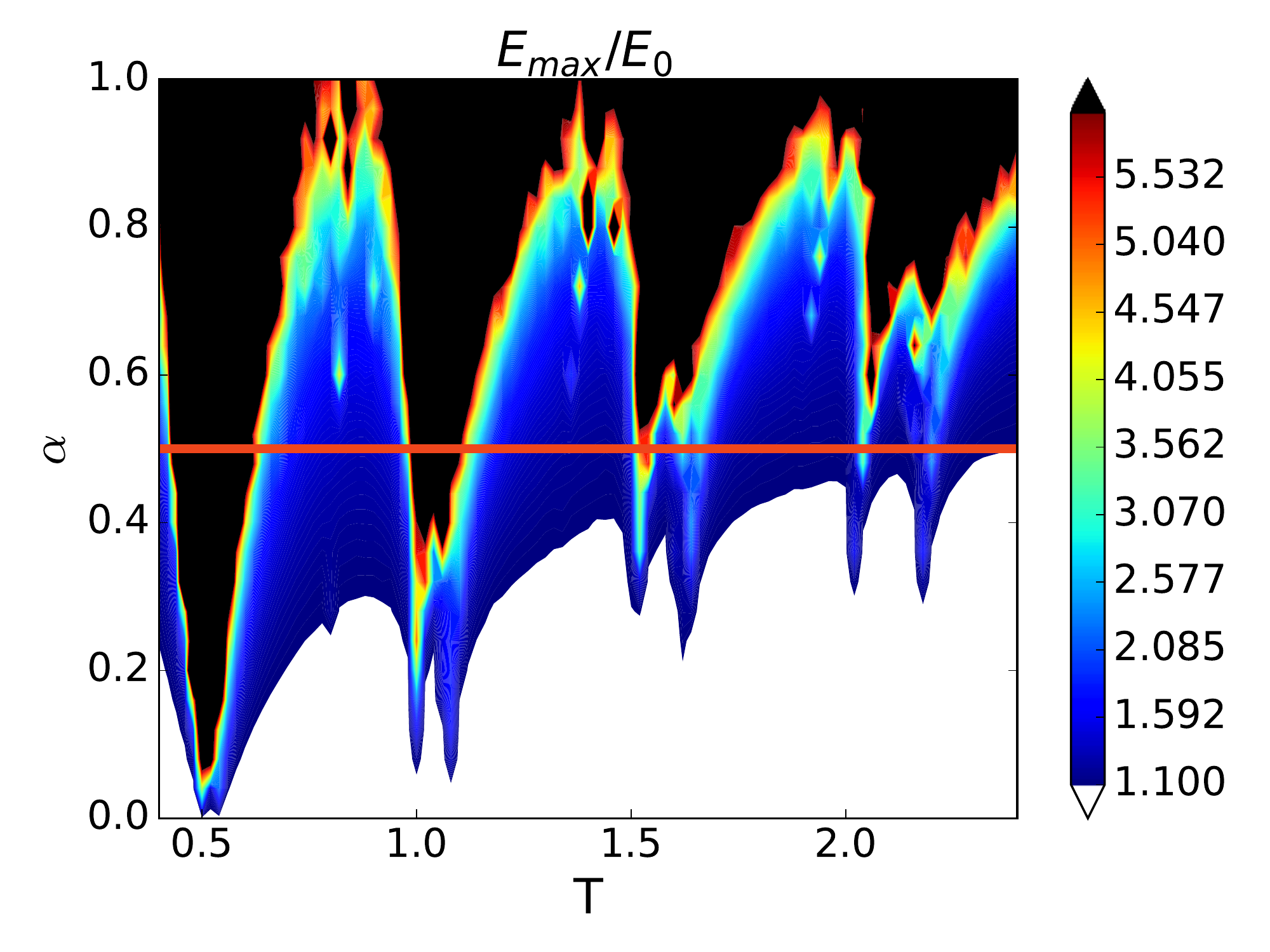}
\caption{\label{Fig:DarmStab}
Stability diagram for the two-body system in the $\alpha$, $T$ parameter space for attractive (left), noninteracting (center), and repulsive (right) cases.
The solid line [green for (a) and red for (b) and (c)] corresponds to $\alpha=0.5$ and is the line we focused on for studying the entanglement entropy.}
\end{figure*}
In Ref.~\cite{Porter:2016yaf}, where no external potential was present, the transfer matrix was approximated for 
small timestep $\tau$ using a symmetric Trotter-Suzuki decomposition as
\beq
\hat{\mathcal T} = e^{-\tau \hat T/2}e^{-\tau \hat V}e^{-\tau \hat T/2} + O(\tau^3).
\eeq
In the present work, we extend the above to include $\hat V_\text{ext}$ by setting 
\beq
\hat{\mathcal T} = e^{-\tau \hat V_\text{ext}/4}e^{-\tau \hat T/2}e^{-\tau \hat V_\text{ext}/4}e^{-\tau \hat V_I}e^{-\tau \hat V_\text{ext}/4}e^{-\tau \hat T/2}e^{-\tau \hat V_\text{ext}/4} + O(\tau^3).
\eeq

With this new transfer matrix, we evolve a trial state in imaginary time to project onto the ground state of the system
by maintaining a constant $\hat V_\text{ext}$.
From that point on, we implement a Wick rotation and set $\tau \to i \delta t$ to study the real-time dynamics of the ground state under
time variations in $\hat V_\text{ext}$ (as specified in the Introduction; see below for more details).

In our calculations, we used a discretized version of the Hamiltonian of Eq.~(\ref{Eq:Hamiltonian}) with the kinetic 
energy piece implemented in momentum space and the interaction and external potential pieces applied in coordinate space. 
Our calculations were carried out in a spatial lattice of $N_x$ points where the lattice spacing was defined as $\ell = 1$ to set the scale for 
every other quantity. The evolution in imaginary time was performed with 
a lattice spacing $\tau = 0.005 \ell^2$, whereas the real-time evolution used a smaller spacing $\delta t = 0.001 \ell^2$.
The dimensionful quantities defining the Hamiltonian of Eq.~(\ref{Eq:Hamiltonian}) are $g$, $\omega_0$ and $T$. From those, 
we form two dimensionless parameters $g^2/\omega_0$ and $\omega T$, which are held constant to their desired 
physical values when taking the continuum limit. To take that limit, we must ensure the following ordering of scales
\beq
\ell \ll 1/\sqrt{\omega_0} \ll  L = N_x \ell,
\eeq
\begin{wrapfigure}{hr}{80mm}
%	\vspace{-100pt}
	\begin{center}
		\includegraphics[width=0.5\columnwidth]{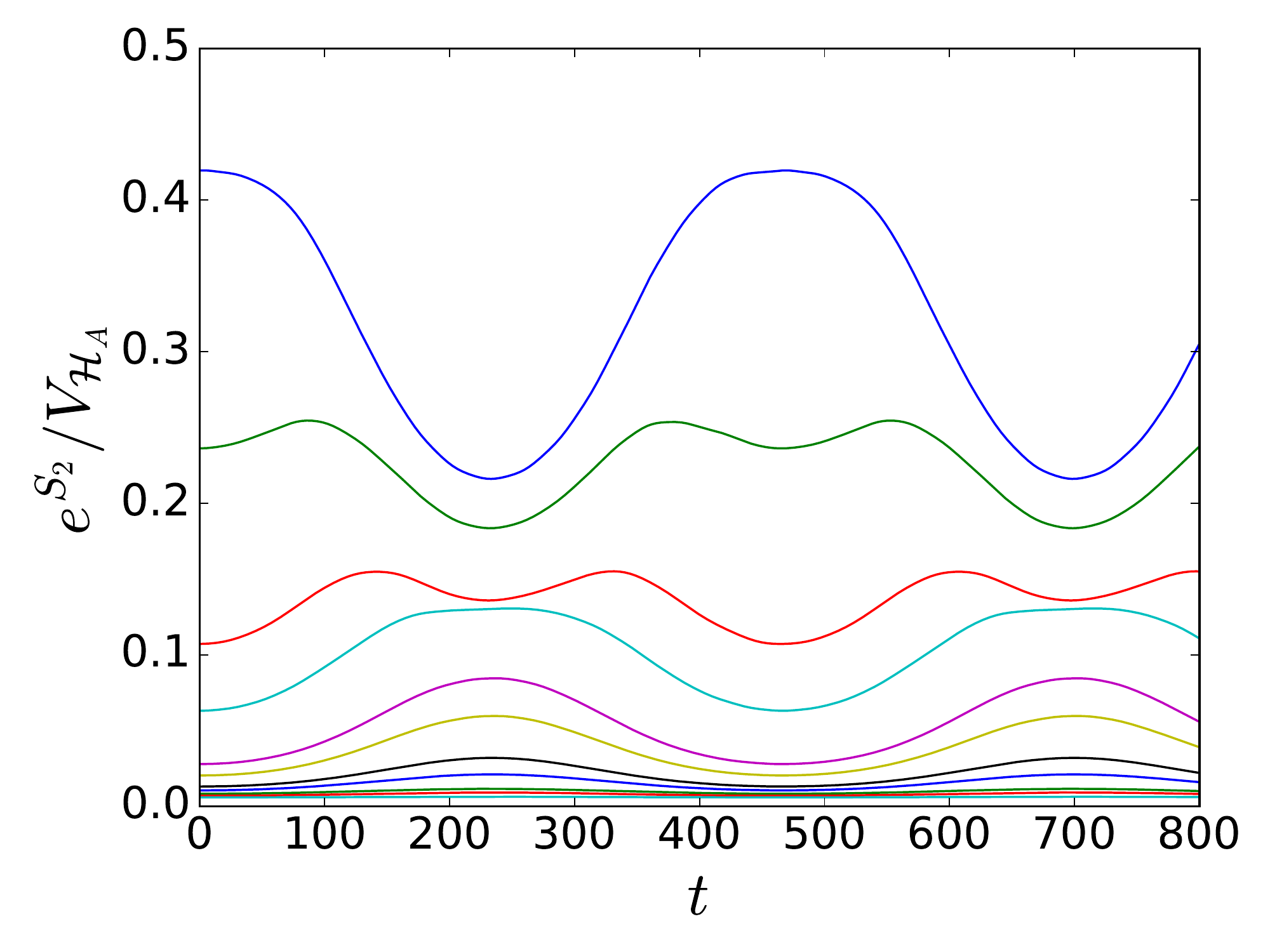}
	\end{center}
	\vspace{-15pt}
\caption{\label{Fig:HilbertFraction} 
Fraction of Hilbert space $\mathcal{H}_A$ required to describe state in subsystem of size $L_A$ ($L_A$ increases from top to bottom).
}
\vspace{-15pt}
\end{wrapfigure}
to ensure that the trapping potential is insensitive to the lattice spacing and volume. In practice, we reduced the numerical value of 
$\omega_0$ until no differences were observed. To keep the physics constant, this required reducing $g^2$ and increasing $T$.
This is consistent with the idea that we must have $g^2 \ll 1/\ell^2$ to ensure that the internal length scales of the two-body interaction 
are insensitive to the lattice spacing. Below, whenever a specific value of $g$ is quoted, it should be understood as corresponding to 
$\omega_0=1$.  We used lattice sizes $N_x = 20,40$ and found the finite-volume effects on the (1+1)- and (2+1)- particle 
systems to be negligible.

It is worth noting that for the one-dimensional systems considered here, other methods such as conventional exact diagonalization
are available. However, the approach presented here does not rely on the dense-matrix form of the quantum operators, 
nor on any particular symmetries of the system, and therefore it is applicable ``as is'' to the two- and three-dimensional 
counterpart systems, and in principle to higher particle number as well (memory constraints permitting; see also our comment below
on attempting the 3-body problem).

%%%%%%%%%%%%%%%%%%%%%%%%%%%%%%%%%%%%%%%%%%%%%%
\section{Results}

%%%%%%%%%%%%%%%%%
\subsection{Crosschecks with results of Ebert et al.}

As a cross-check for correctness of our lattice approach, we reproduced the results of Ref.~\cite{Ebert:2015gko}, comparing the two methods for 
the following quantities:

\begin{itemize}

\item[\textbf{(1)}]$\ $ the root-mean-square separation of particles under an exponential driving potential with finite cutoff, which 
show excellent agreement with Ref.~\cite{Ebert:2015gko}, as shown at left in Fig.~\ref{Fig:DarmRMS};

\item[\textbf{(2)}]$\ $ the energy in a harmonic trap external potential with periodic frequency changes for interacting and noninteracting cases (right of Fig.~\ref{Fig:DarmRMS}), which agrees with the corresponding plot of Ref.~\cite{Ebert:2015gko} once the center-of-mass energy is taken into account;

\item[\textbf{(3)}]$\ $ the stability of the system under variation of the harmonic trap parameters $\alpha$ and $T$ (Fig.~\ref{Fig:DarmStab}).
As our approach did not involve an analytical solution, which would clearly indicate regions of instability, we used the maximum energy 
$E_{\text{max}}/E_0$ (average energy for the attractive case) attained over a trajectory to provide a numerical indicator for resonant behavior. 
Our results are quantitatively consistent with those of Ref.~\cite{Ebert:2015gko}.

\end{itemize}

%%%%%%%%%%%%%%%%%%%%%%%%%%%%%%%%%%
\subsection{Local thermal behavior}

The trajectory represented in Fig.~\ref{Fig:HilbertFraction} is the same as that in Fig.~\ref{Fig:DarmRMS} (right). As described in Ref.~\cite{Ebert:2015gko}, the driving frequency is tuned to the gap between the ground and second excited states, such that at $t\approx 225$ and $t\approx 700$, the system is nearly entirely in the excited state, while at $t=0$ and $t\approx 480$, it is in the ground state. Viewed from top to bottom, it is evident that the fraction of the Hilbert space required to described the state decreases as the subsystem size increases, approaching $1/V_{\mathcal{H}_A}$ as $L_A$ approaches the full system size, $L$. As the system is excited,
the smallest subsystem's (topmost curve, blue) Hilbert space occupation decreases because particles are driven away from the center of the trap. For larger subsystems, however, the opposite occurs because prior to the excitation, the particles were mostly contained within the boundaries of $L_A$. Thus, we see that the size of the subsystem observed relative to its exterior can cause a large variation in the description of the quantum state. 
Additionally, we see that the local basis of very small subsystems is inefficient at describing states in the larger basis of the full system, leading to apparent thermal behavior in the subsystem. While in agreement with Ref.~\cite{Kaufman794}, we extend that work by observing the system as a function of time as the external driving continues.

%%%%%%%%%%%%%%%%%%%%%%%%%%%%%%%%%%
\subsection{Entanglement entropy}

\begin{figure*}[h]
\centering\captionsetup{format=plain,justification=centerlast}
	\includegraphics[width=130.0pt]{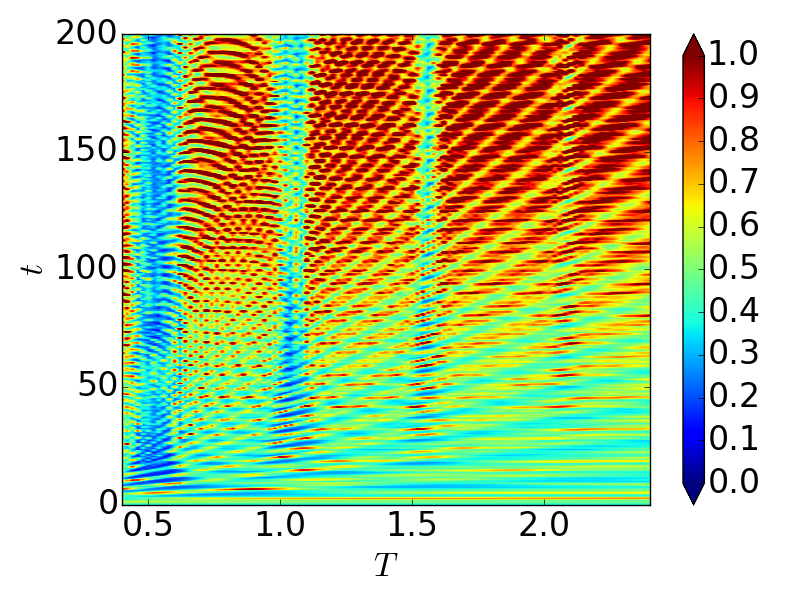}
	\includegraphics[width=130.0pt]{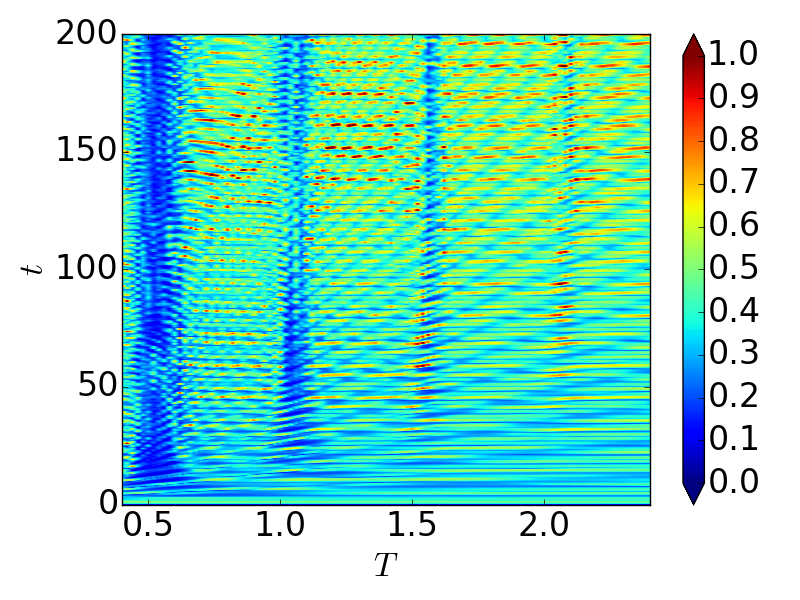}
	\includegraphics[width=130.0pt]{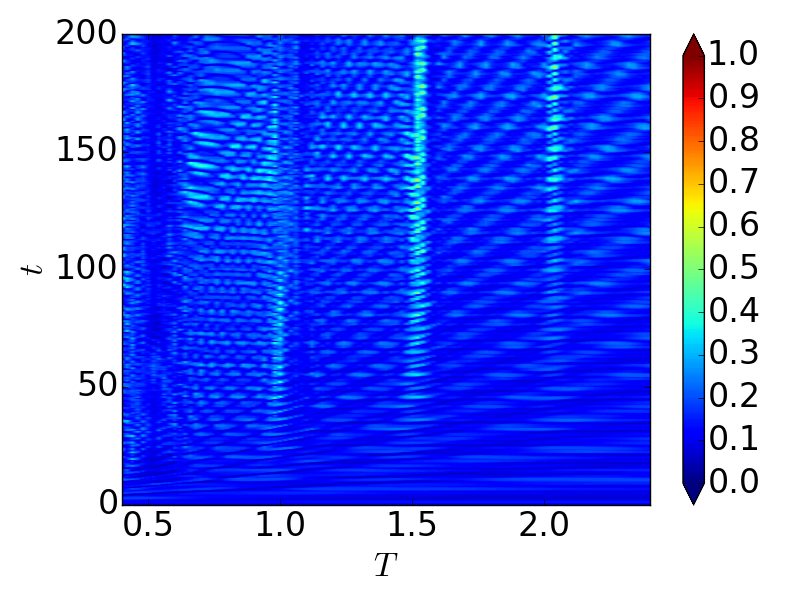}
	\includegraphics[width=130.0pt]{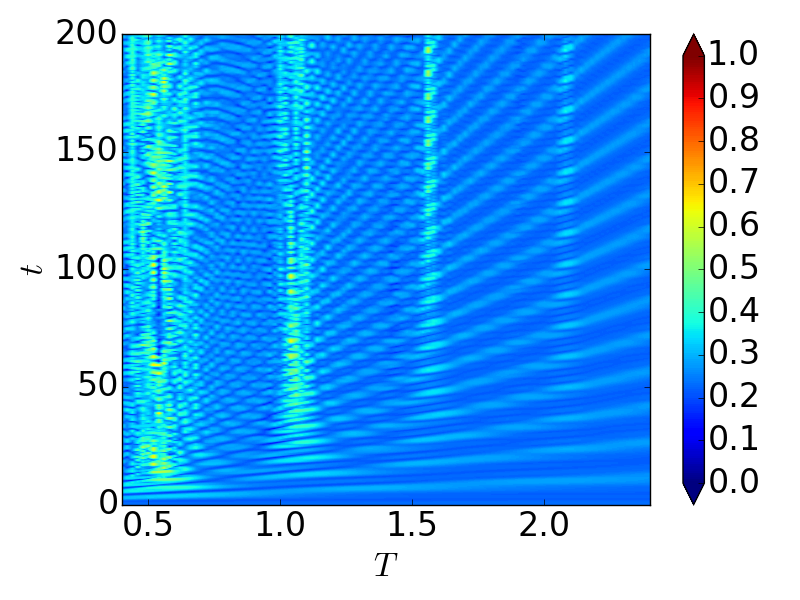}
	\includegraphics[width=130.0pt]{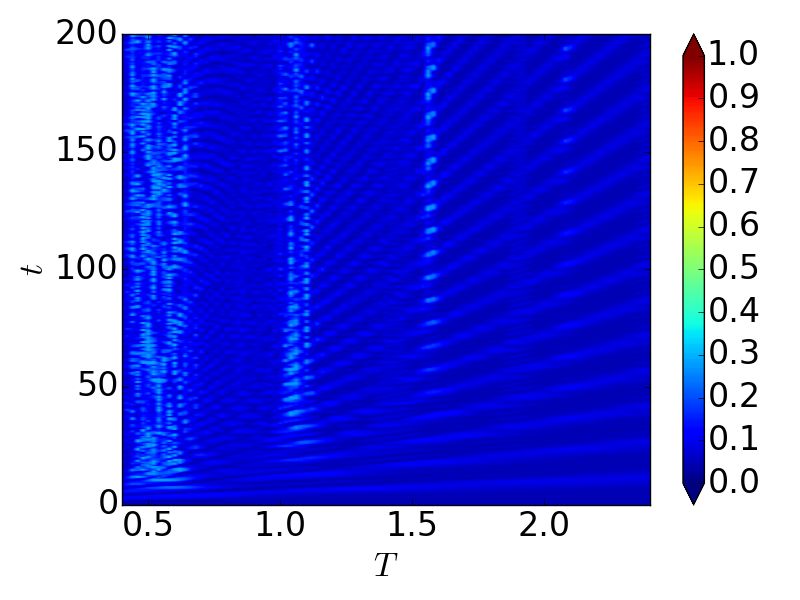}
	\includegraphics[width=130.0pt]{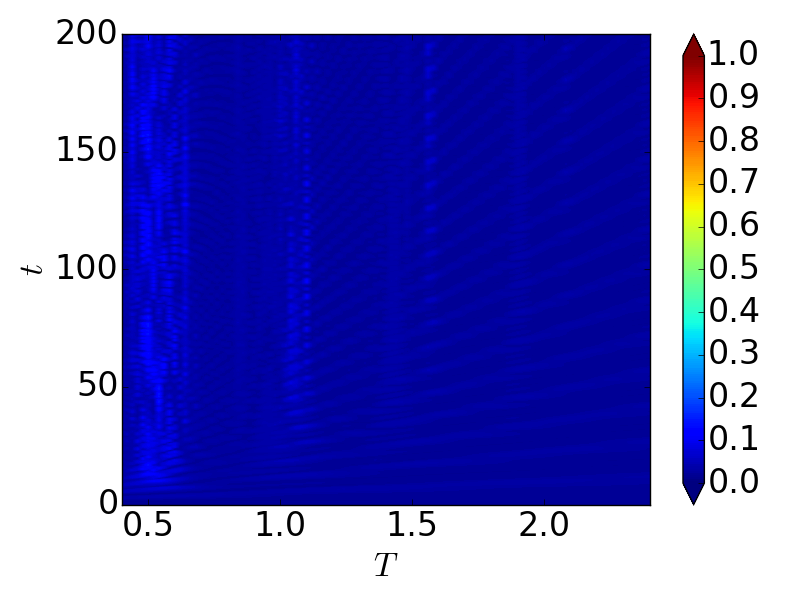}
	\includegraphics[width=130.0pt]{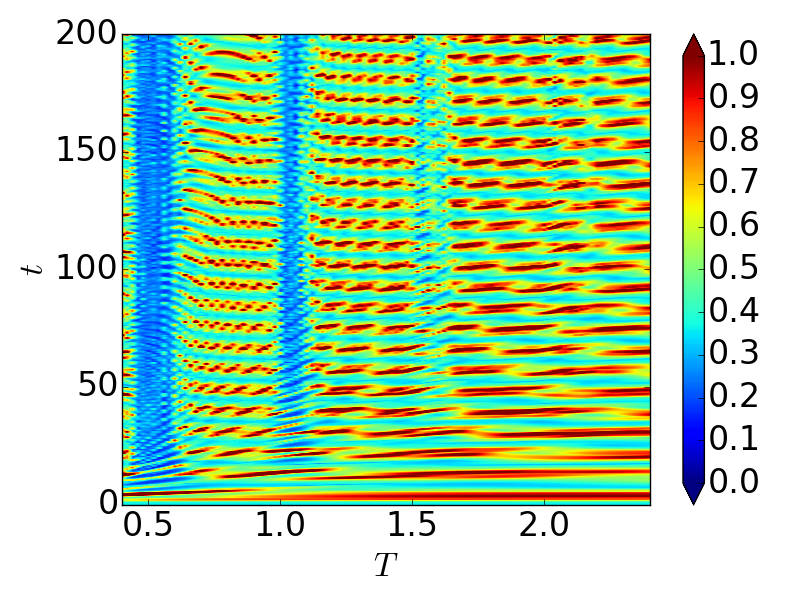}
	\includegraphics[width=130.0pt]{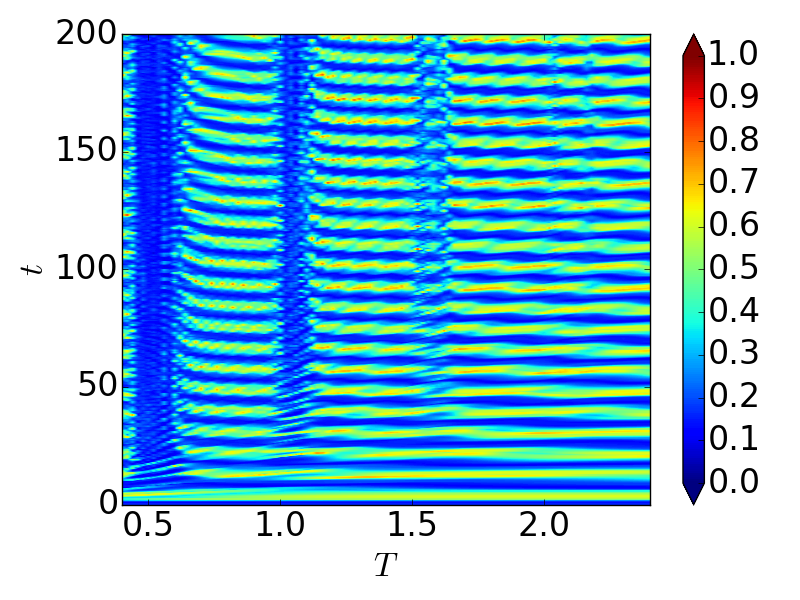}
	\includegraphics[width=130.0pt]{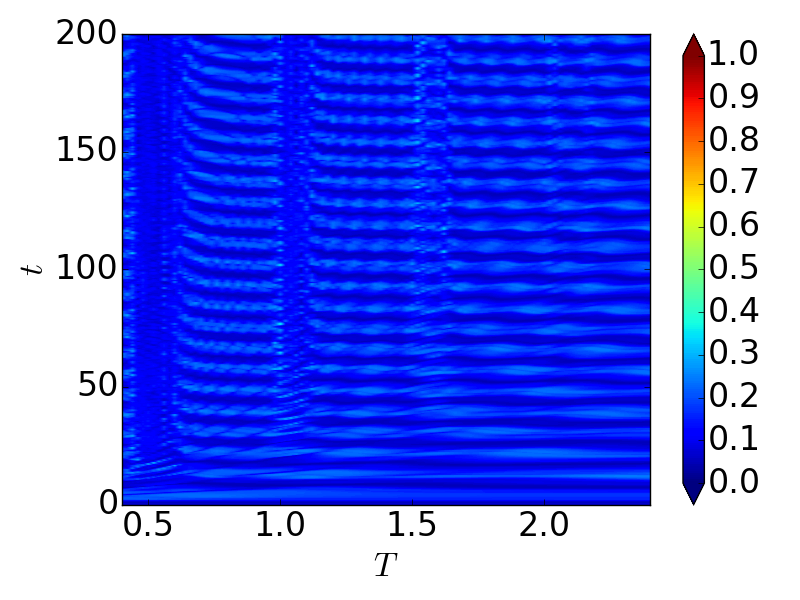}
\caption{\label{Fig:EETtg0}
Time evolution of the second R\'enyi entanglement entropy $S_2$ plotted as $\exp(S_2)/V_\mathcal{H_A}$
(see text for details) for subsystem sizes $L_A = 2$ (left), $3$ (center), and $4$ (right) sites 
(concentric with the trap center) for {\bf attractive} (first row, $g = 2.0$), {\bf noninteracting} (second row, $g=0$), and 
{\bf repulsively interacting} (third row, $g=-2.0$) 
systems of 2+1 particles. The data is shown as a function of time $t$ and driving parameter $T$
crossing the resonances present at $\alpha = 0.5$ around $T\simeq 0.5$, $1.0$, and $1.5$ (see stability diagrams).}
\end{figure*}

In this section we present 
the calculation of the entanglement entropy across a parametric resonance, focusing for definiteness
on the case $\alpha = 0.5$ (see Fig.~\ref{Fig:DarmStab}).
In the absence of interactions, the trap oscillations drive the system across a parametric resonance which,
for $\alpha = 0.5$, is present around $T\simeq 0.5$, $1.0$, and $1.5$ (and traces of another one can be seen in 
Fig.~\ref{Fig:DarmStab} for $T\simeq 2.1$). In Fig.~\ref{Fig:EETtg0} we show the impact of the driving potential 
on the time evolution of the second R\'enyi entanglement entropy $S_2$. While it is a non-trivial problem to understand
the behavior of $S_2$ analytically (because it is a complicated quantity that varies with the subsystem size $L_A$),
certain features can be gleaned which, as we will see below, are a clear imprints of the interactions, while
other features are essentially unchanged relative to the noninteracting case.

Specifically, we see that the resonance yields a seemingly chaotic signal when plotted in the $t,T$ plane (to be contrasted 
with the more regular structure observed away from the resonance in those plots) and changes the entanglement 
entropy by a large factor. In Fig.~\ref{Fig:EETtg0} we show the second R\'enyi entanglement entropy $S_2$ plotted as 
$\exp(S_2)/V_{\mathcal{H}_A}$, where $V_{\mathcal{H}_A}$ is the number of states in the Hilbert space where $S_2$ is 
computed (which depends on the  subsystem size $L_A$).
Since this work started, we have noticed that the time-evolution operator did not remain exactly unitary for interacting (2+1)-particle systems. 
While this is likely a numerical issue, we have yet to determine the exact cause. This non-conservation of probability may very well account 
for the unexpected increase in entropy seen in Fig.~\ref{Fig:EETtg0} over time, as the evolution becomes essentially random.

%%%%%%%%%%%%%%%%%%%%%%%%%%%%%%%%%%%%%%%%%%%%%%
\section{Summary and Conclusions}

In this work we studied the time evolution of systems of 1+1 and 2+1 spin-1/2 particles in a one-dimensional
time-varying external trapping potential. Expanding on the work of Ref.~\cite{Ebert:2015gko}, which we reproduced as a 
crosscheck for the (1+1)-particle case, we determined the impact on the entanglement entropy of driving the 
system through a resonance. It is worth noting that we used very different techniques from Ref.~\cite{Ebert:2015gko}: 
we put the particles on a spatial lattice and carried out the time evolution by first projecting onto the ground state 
(thus preparing the system) using imaginary-time evolution (i.e. applying the transfer matrix), followed by real-time 
evolution to study the dynamics.

The objective of Ref.~\cite{Ebert:2015gko} was to investigate the interaction effects on the two-body problem by considering
two different scenarios: a rapid trap change and a periodic trap change. While the former behaved essentially as the non-interacting case,
the latter, which is the main one we considered here, can enhance small fluctuations by virtue of parametric resonance. Such a situation 
is interesting for several reasons: small interaction effects may have a large impact on the overall behavior of the system;
the driving protocol is similar to the quantum-control protocol used in experiments; and resonances can be used for heating or cooling.

To add insight into this problem, and inspired by the work of Ref.~\cite{Kaufman794}, we examined the emergence of local thermal 
behavior and showed how it varied over time under the influence of external driving. To this end, we put forward a non-stochastic
lattice technique that only incurs uncertainties in lattice spacing. As expected, the R\'enyi entanglement entropy showed thermal 
behavior diminishing as the subsystem size grew to the full system. This work is therefore a very first step towards building a more
robust technique that can address larger few-body problems (current memory limitations should allow for at least 10 particles in
three spatial dimensions, but using supercomputers and aggressive optimization) and a much wider range of observables and
time-varying situations.

%%%%%%%%%%%%%%%%%%%%%%%%%%%%%%%%%%%%%%%%%%%%%
{
This material is based upon work supported by the
National Science Foundation under Grant No.
PHY{1452635} (Computational Physics Program).
}

\clearpage
\bibliography{lattice2017}

%%%%%%%%%%%%%%%%%%%%%%%%%%%%%%%%%%%%%%%%%%%%%%%%%%%%%%%%%%%%%%%%%%%%%%%%%%%%%
\end{document}